\documentclass[manuscript]{aastex}

\def\mb#1{\mbox{\boldmath $#1$}}
\slugcomment{Not to appear in Nonlearned J., 45.}

\shorttitle{Dynamo mechanism in galactic gaseous disk}
\shortauthors{Machida et al.}

\begin{document}

\title{DYNAMO ACTIVITIES DRIVEN BY MAGNETO-ROTATIONAL INSTABILITY 
AND PARKER INSTABILITY IN GALACTIC GASEOUS DISK}

\author{MACHIDA Mami\altaffilmark{1}, NAKAMURA Kenji, E.\altaffilmark{2}, 
KUDOH Takahiro\altaffilmark{3}, AKAHORI Takuya\altaffilmark{4,5}, SOFUE Yoshiaki\altaffilmark{6,7}}
\and

\author{MATSUMOTO Ryoji\altaffilmark{8}}

\altaffiltext{1}{Department of Physics, Faculty of Sciences, Kyushu University,
6-10-1, Hakozaki, Higashi-ku, Fukuoka, 812-8581, Japan: mami@phys.kyushu-u.ac.jp}
\altaffiltext{2}{Department of Mechanical Engineering, Kyushu Sangyo University, 
2-3-1, Matsukadai, Higashi-ku, Fukuoka, 813-8503, Japan}
\altaffiltext{3}{National Astronomical Observatory of Japan, Mitaka, Tokyo 181-8588, Japan}
\altaffiltext{4}{Korea Astronomy and Space Science Institute, Daedeokdaero 776, Yuseong-Gu, Daejeon 305-348, 
Republic of Korea}
\altaffiltext{5}{Sydney Institute for Astronomy, School of Physics, The University of Sydney, 
44 Rosehill Street, Redfern, NSW 2016, Australia}
\altaffiltext{6}{Institute of Astronomy, University of Tokyo, Mitaka 181-8588, Tokyo, Japan}
\altaffiltext{7}{Department of Physics, Meisei University, Hino 191-8506, Tokyo, Japan} 
\altaffiltext{8}{Department of Physics, Graduate School of Science, Chiba University,
1-33, Yayoi-cho, Inage-ku, Chiba, 263-8522, Japan}

\begin{abstract}
 We carried out global three-dimensional magneto-hydrodynamic 
simulations of dynamo activities in galactic gaseous disks 
without assuming equatorial symmetry.
Numerical results indicate the growth of 
azimuthal magnetic fields non-symmetric to the equatorial plane.
As magneto-rotational instability (MRI) grows,  
the mean strength of magnetic fields is amplified until the magnetic pressure
becomes as large as 10\% of the gas pressure. 
When the local plasma $\beta$ ($ = p_{\rm gas}/p_{\rm mag}$) becomes less than 5 
near the disk surface, 
magnetic flux escapes from the disk by Parker instability 
within one rotation period of the disk. 
The buoyant escape of coherent magnetic fields drives dynamo 
activities by generating disk magnetic fields with opposite polarity 
to satisfy the magnetic flux conservation.
The flotation of the azimuthal magnetic flux from the disk and 
the subsequent amplification of 
disk magnetic field by MRI 
drive quasi-periodic reversal of azimuthal magnetic fields 
in timescale of 10 rotation period.
Since the rotation speed decreases with radius, 
the interval between the reversal of azimuthal magnetic fields 
increases with radius.
The rotation measure  computed from the numerical results 
shows symmetry corresponding to a dipole field. 

\end{abstract}


\section{Introduction}
The origin and nature of magnetic fields in spiral galaxies 
is an open question. 
Magnetic fields in spiral galaxies have been intensively studied through 
observations of linear polarization of synchrotron radiation 
\citep{sof1986, bec1996}. 
The typical magnetic field strength is a few $\mu {\rm G}$ 
in the Milky Way near the Sun,  and in spiral galaxies.  
The mean ratio of gas pressure to magnetic pressure is close to unity \citep{ran1989, cro2010}, 
suggesting that magnetic fields in spiral galaxies are strong enough to 
affect the motion of the interstellar matter in the gaseous disk.

Since the galactic gas disk rotates differentially, 
the magnetic field is amplified faster 
near the galactic center than in the outer disk,  
and the magnetic field strength can exceed $0.1 {\rm mG}$ 
in the galactic center. 
An observational evidence of such strong magnetic fields 
has been obtained in 
the radio arc near the galactic center, where the 
field strength of about $1 {\rm mG}$ is estimated from synchrotron radiation. 
Many threads perpendicular to the galactic 
plane are observed indicating strong vertical field 
\citep{yus1984, mor1990}. 
On the other hand, observations of near infrared polarization suggested that 
horizontal fields of $0.1 - 1 {\rm mG}$ 
are dominant inside the disk 
\citep{nov2000, nis2009}. 

\citet{tay2009} presented results of all sky distribution of rotation measure (RM),
which indicate that the magnetic fields along line-of-sights are roughly 
consistent with the dipole-like models in which the azimuthal magnetic fields 
are anti-symmetric with respect to the equatorial plane, 
rather than quadrupole-like models in which the azimuthal magnetic fields show 
symmetry with respect to the equatorial plane. 
They also showed that the RM distribution has small scale variations, 
which would be linked to turbulent structure of the gaseous disk. 
\citet{han2002} showed a view of RM distributions from the 
North Galactic pole using the results of pulsar polarization 
measurements. 
They indicated that the magnetic field direction changes between the 
spiral arms \citep{han2002, han2007}.

The magnetic fields have to be maintained for $10^9 - 10^{10}$ years, 
because most of  spiral galaxies 
have magnetic fields with similar strength \citep{bec2011}.
\citet{par1971} considered the self-sustaining mechanism of magnetic fields 
called the dynamo activity. 
In the framework of the kinematic dynamo, where the back-reaction by enhanced 
magnetic fields is ignored, 
\citet{fuj1987} and \citet{saw1987} obtained the global structure of galactic magnetic fields. 
On the other hand, \citet{bal1991} 
pointed out importance of 
the magneto-rotational instability (MRI) in differentially rotating system,  
and showed that the back-reaction of the enhanced magnetic fields 
cannot be ignored. 
Since galactic gaseous disk is a differentially rotating system, 
magneto-rotational instability grows \citep{bal1991}.

In this paper, we present the results of global three-dimensional 
magneto-hydrodynamic simulations of  galactic gas disks 
assuming an axisymmetric gravitational potential and 
study the dynamo activities of the galactic gas disk. 

In order
to take into account the back reaction of magnetic fields 
on the dynamics of gas disk, 
we have to solve the magneto-hydrodynamic equations in which 
the induction equation is coupled with the equation of motion. 
This approach is called the dynamical dynamo. 
Nishikori, Machida \& Matsumoto (2006) presented results of three dimensional 
magneto-hydrodynamic (MHD) simulations of galactic gaseous disks. 
They 
showed that, even when the initial magnetic field is weak, 
the dynamical dynamo amplifies the magnetic fields 
and can attain the current field strength observed in galaxies. 
They also  
showed that the direction of mean azimuthal magnetic fields reverses 
quasi-periodically, and 
pointed out that 
reversal of magnetic field is driven by the 
coupling of MRI and Parker instability. 
When the magnetic field is amplified by MRI, 
the time scale of buoyant escape of magnetic flux 
due to Parker instability becomes comparable to 
the growth time of MRI, 
and limits the strength of magnetic fields inside the disk.

Since \citet{nis2006} assumed 
symmetric boundary condition at the equatorial plane of the disk, 
they could not simulate the growth of anti-symmetric mode of  
Parker instability, in which magnetic fields cross the 
equatorial plane \citep{hor1988}. 
The present paper is aimed at obtaining a realistic, quantitative model 
of the global magnetic field of the Milky Way Galaxy based on 
numerical 3D MHD simulations of magnetized galactic gaseous disk. 
In section 2, we show the initial setting of the simulation. 
We present the results of numerical simulations in section 3. 
The result will be used to compute the distribution of RM on the 
whole sky in section 4.

\section{Numerical Models}
\subsection{Basic equations}
We solved the resistive MHD equations
by using modified Lax-Wendroff scheme with artificial viscosity 
(Rubin \& Burstein 1967, Richtmyer \& Morton 1967). 

\begin{equation}
\frac{\partial \rho}{\partial t} + \nabla \cdot{ \rho \mb{v}} = 0
\end{equation}
\begin{equation}
\frac{\partial \mb{v}}{\partial t} + (\mb{v} \cdot \nabla)\mb{v} = 
-\frac{1}{\rho} \nabla P - \nabla \phi + 
\frac{1}{4 \pi \rho} (\nabla \times \mb{B}) \times \mb{B} 
\end{equation}
\begin{equation}
\frac{\partial}{\partial t} \left(
\frac{1}{2}\rho v^2 + \frac{B^2}{8 \pi} + \frac{P}{\gamma -1}
\right) 
+
\nabla \cdot \left[
\left(
\frac{1}{2}\rho \mb{v}^2 + \frac{\gamma P}{\gamma-1}
\right) \mb{v} + \frac{c}{4 \pi} \mb{E} \times \mb{B}
\right] 
= - \rho \mb{v} \nabla \phi
\end{equation}
\begin{equation}
\frac{\partial \mb{B}}{\partial t} = 
\nabla \times (\mb{v} \times \mb{B} - \eta \nabla \times \mb{B})
\end{equation}
where $\rho$, $P$, $\mb{v}$, $\mb{B}$, and $\gamma$ are 
the density, gas pressure, velocity, magnetic field, the specific heat ratio, respectively,  
and $\phi$ is gravitational potential proposed by Miyamoto \& Nagai (1975). 
The electric field $\mb{E}$ is related to magnetic field $\mb{B}$ by 
Ohm's law. 
We assume the anomolous resistivity $\eta$ 
\begin{equation}
\eta = \left\{ 
\begin{array}{ll}
\eta_0 (v_{\rm d}/v_{\rm c} - 1)^2, & v_{\rm d} \ge v_{\rm c}, \\
0,& v_{\rm d} \le v_{\rm c},
\end{array}
\right.
\end{equation}
which sets in when the electron-ion drift speed $v_{\rm d} = j/\rho$, 
where $j$ is the current density, exceeds the critical velocity $v_{\rm c}$.

\subsection{Simulation Setup}
We assumed an equilibrium torus threaded by weak toroidal magnetic fields 
\citep{oka1989}. 
The torus is embedded in a hot, non-rotating spherical corona. 
We assume that the torus has constant specific angular momentum $L$ at $\varpi = \varpi_{\rm b}$  
and assume polytropic equation of state $P = K \rho^{\gamma}$, 
where $K$ is constant. 
The density distribution is same as that for equation (8) in \citet{nis2006}.

We consider a gas disk composed of one-component interstellar gas. 
Radiative cooling and self-gravity of gas are ignored in this paper. 

We adopt a cylindrical coordinate system $(\varpi, \varphi, z)$. 
The units of length and velocity in this paper are $\varpi_0 =1 \hspace{1mm} \rm{kpc}$ and 
$v_0=(GM/\varpi_0)^{1/2} = 207 \hspace{1mm} \rm{km/s}$, respectively, 
where $M =10^{10} M_{\odot}$. 
The unit time is $t_0 = \varpi_0/v_0$.
Other numerical units are listed in Table 1. 

\begin{table}
\begin{center}
\caption{Units of Physical quantities.}
\begin{tabular}[t]{ccc} \hline \hline
Physical quantity & Symbol & Numerical unit \\ \hline
Length & $\varpi_0$ & 1 kpc \\
Velocity & $v_0$ & 207 km/s \\
Time & $t_0$ & 4.8 $\times 10^6$  yr\\
Density & $\rho_0$ & 1.6 $\times 10^{-24} {\rm g/cm^3}$ \\
Temperature & $T_0$ & 5.15 $\times 10^6$ K \\
Magnetic field & $B_0$ & 26 $\mu$ G \\ \hline
\end{tabular}
\end{center}
\end{table}

The numbers of grid points we used  are $(N_{\varpi}, N_{\varphi}, N_z) = (250,128,640)$. 
The grid size is $\Delta \varpi/\varpi_0 = 0.05$, $\Delta z/\varpi_0 = 0.01$ for 
$0 \leq \varpi/\varpi_0 \leq 6.0$, $|z|/\varpi_0 \leq 2.0$, and otherwise they 
increase with $\varpi$ and $z$, respectively. 
The grid size in azimuthal direction is $\Delta \varphi = 2 \pi/128$. 
The outer boundaries at $\varpi=56 \varpi_0 $ and $|z|=10 \varpi_0$ are 
free boundaries where waves can be  transmitted. 
We applied periodic boundary conditions in the azimuthal direction. 
An inner absorbing boundary condition is imposed at 
$r = \sqrt{\varpi^2 + z^2} = r_{\rm in} = 0.8 \varpi_0$, 
since the gas accreted to the central region of the galaxy 
will be converted to stars or swallowed by the black hole. 
To  illuminate the effect of boundary condition at the 
equatorial plane, 
we calculate two models: one is model PSYM which imposes 
symmetric boundary condition at the equatorial plane.
This model is identical with that in  \citet{nis2006}. 
The other is model ASYM,  
in which we include the equatorial plane inside the 
simulation region, and imposed no boundary condition 
at the equatorial plane.    
In this paper, we adopted model parameters $\varpi_{\rm b} = 10 \varpi_0$, 
$\beta_{\rm b} = 100$ at $(\varpi, z)=(\varpi_{\rm b},0)$, $\gamma = 5/3$, $v_{\rm c} = 100 v_0$, 
and $\eta_0=0.1 v_0 \varpi_0$. 
The thermal energy of the torus is parameterized by 
$E_{\rm th} = c_{\rm sb}^2/(\gamma v_0^2)= K \rho_{\rm b}^{\gamma -1}/v_0^2 = 0.05$ 
where $c_{\rm sb}$ and $\rho_{\rm b}$ are the sound speed and density at 
$(\varpi, z)=(\varpi_{\rm b}, 0)$, respectively. 
The initial temperature of the torus is $T \sim 2 \times 10^5$K. 
This mildly hot plasma mimics the mixture of hot ($T \sim 10^6$K) 
and warm ($T \sim 10^4$K) components of the interstellar matter, 
which occupy large volume of the galactic disk.

\section{Numerical results}

\subsection{Time evolution of the galactic gaseous disk}

\begin{figure}
\plotone{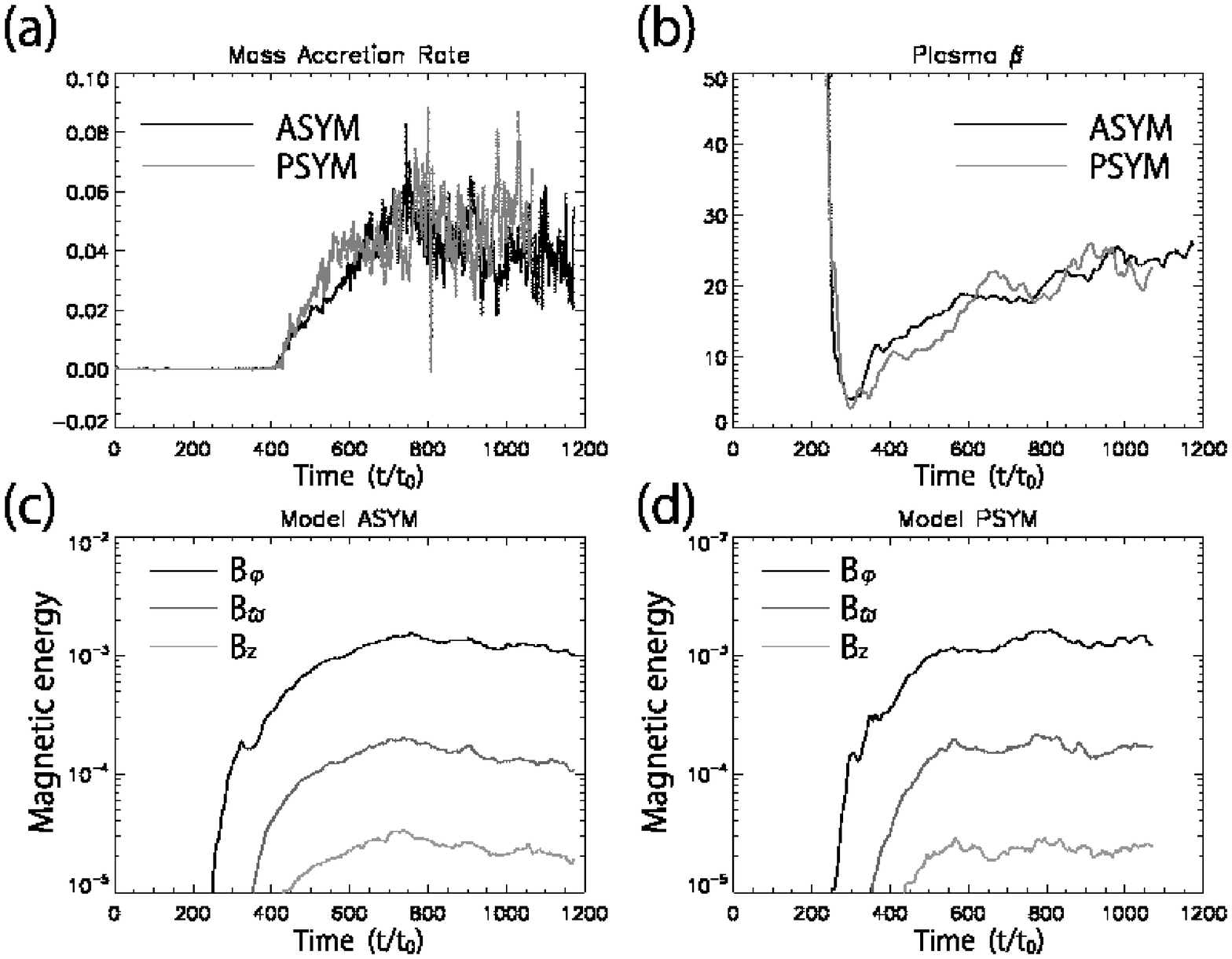}
\caption{(a) Time evolution of the mass accretion rate at $\varpi/\varpi_0 = 2.5$. 
Black curve shows model ASYM and gray curve shows model PSYM. 
(b) Time evolution of the plasma $\beta$ averaged in the region where 
$2 < \varpi/\varpi_0 <5$, $|z|/\varpi_0 < 1$, and $0 \leq \varphi \leq 2 \pi$. 
(c) Time evolution of the spatially averaged magnetic energy of model ASYM. 
Black, dark gray and gray denotes azimuthal component, radial component 
and vertical component, respectively. 
(d) Time evolution of the spatially averaged magnetic energy of model PSYM. 
Colors are same as in (c). The averaged region of the magnetic energy is 
same as (b).
\label{fig1}}
\end{figure}

Figure \ref{fig1}a shows the time evolution of the mass accretion rate measured at  
$\varpi/\varpi_0=2.5$. 
After $t/t_0 \sim 200$, the gas starts to inflow and 
the mass accretion rate increases linearly. 
Subsequently ($t/t_0 > 600 $) the accretion rate saturates and becomes roughly constant. 
The evolution of the mass accretion rate in model ASYM (black) is similar 
to that in model PSYM (gray), 
where the mass accretion rate for model PSYM is doubled 
since the model PSYM 
includes only the region above the equatorial plane.

Figures \ref{fig1}b, \ref{fig1}c and \ref{fig1}d show the time evolution of 
the plasma $\beta$ and magnetic energy of each component 
averaged in the region where $2 \leq \varpi/\varpi_0 \leq 5$, 
$|z|/\varpi_0 \leq 1$, and $0 \leq \varphi \leq 2 \pi$,  respectively.  
It is clear that plasma $\beta$ decreases in the linear stage as the magnetic energy increases. 
In both models, the averaged magnetic energy first increases exponentially. 
After that, it saturates and becomes roughly constant. 
Time evolution of magnetic fields is similar between 
model ASYM and model PSYM.
Time evolutions of the growth and saturation in the mass accretion rate 
are quite similar to those of the averaged magnetic energy.  
It means that the magnetic turbulence driven by MRI is the main cause of 
the angular momentum transport which drives mass accretion. 
In the remaining part of this section, we discuss the results of model ASYM.

\begin{figure}
\plotone{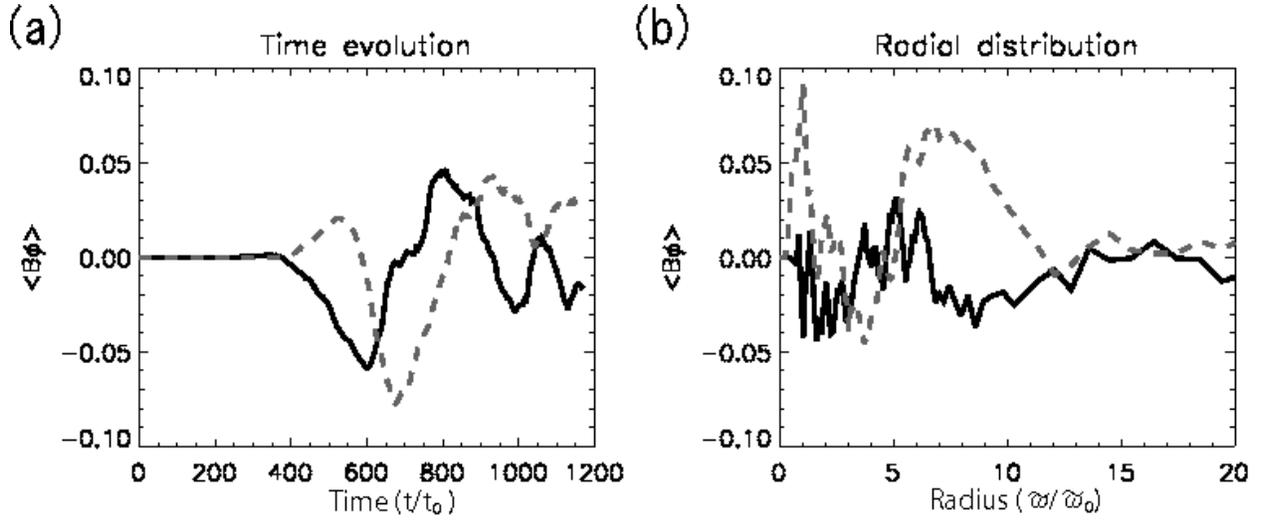}
\caption{Evolutions of the mean azimuthal magnetic fields 
averaged in the region where $5 <\varpi/\varpi_0 < 6$, 
$0 \leq \varphi < 2 \pi$, and $0<z/\varpi_0 < 1$ (black), 
and  in the region where $5 <\varpi/\varpi_0 < 6$, 
$0 \leq \varphi < 2 \pi$, and $1<z/\varpi_0 < 3$ (gray). 
(a) and (b) show time evolution and radial distribution, 
respectively.
\label{fig2}}
\end{figure}

In order to check  the magnetic fields structure, 
we analyze the time evolution of mean azimuthal magnetic fields. 
The mean fields are computed by the same method as \citet{nis2006}.
Figure \ref{fig2}a shows the time evolution of mean azimuthal magnetic fields.
Black curve shows $B_{\varphi}$ averaged in the region where $5 < \varpi/\varpi_0 < 6$, 
$0 \leq \varphi \leq 2 \pi$, and $0 < z/\varpi_0 < 1$, and 
gray curve shows $B_{\varphi}$ averaged in the region where $5 < \varpi/\varpi_0 < 6$, 
$0 \leq \varphi \leq 2 \pi$, and $1 < z/\varpi_0 < 3$. 
Black and gray curves correspond to the disk region and halo region, respectively.
The azimuthal magnetic fields reverse their direction with timescale 
$t \sim 300 t_0 \sim 1.5 {\rm Gyr}$. 
This timescale is comparable to that of the buoyant rise of azimuthal magnetic fields.
The halo fields also reverse their directions with the same 
time scale as the disk component. 

The radial distribution of the mean azimuthal magnetic fields at $t = 1000 t_0$
are shown in figure \ref{fig2}b.
Black and gray curves show the averages in the disk 
($ 0 \leq \varphi \leq 2 \pi$ and $0 < z/\varpi_0 < 1$), and 
in the halo  ($ 0 \leq \varphi \leq 2 \pi$ and $1 < z/\varpi_0 < 3$), respectively. 
Since the magnetic fields in the disk becomes turbulent,  
the direction of azimuthal magnetic fields changes frequently near the equatorial plane. 
On the other hand, long wavelength magnetic loops are formed in the halo region.

Figure \ref{fig3}a shows the radial distribution of the specific angular momentum at 
equatorial plane.
Although the rotation is assumed only inside initial torus (gray curve 
in figure \ref{fig3}a), 
torus deforms its shape from a torus to a disk by redistributing angular momentum 
(black curve in figure \ref{fig3}a). 
Since the outflow created by the MHD driven dynamo transports angular momentum 
toward the halo, the halo begins to rotate differentially, 
as can be seen in iso-contours of the azimuthally averaged rotation speed 
(figure \ref{fig3}b).  
Figure \ref{fig3}a and \ref{fig3}b show that 
the azimuthal velocity $v_{\varphi}$ is almost constant 
and close to $v_0$ in the equatorial plane. 
The rotation period at $\varpi/\varpi_0 = a$ is 
$P_{\rm rot} \sim 2 \pi a (\varpi_0/ v_0) = 2 \pi a t_0$. 

\subsection{Evolution of the magnetic field structure}

We show the 3D structure of mean magnetic fields in figure \ref{fig4}.
The colored surfaces represent iso-density surfaces, 
and the curves show magnetic field lines. 
Light brown curves show field lines passing through the plane at 
$z/\varpi_0 = 5$ and $\varpi/\varpi_0 < 1$.  
The vertical magnetic fields around the rotation axis 
are produced by the magnetic pressure driven outflow near the galactic center. 
Color on the curves depicts the direction of azimuthal magnetic fields. 
Red curves show positive direction of azimuthal magnetic fields 
(counter clockwise direction) and 
blue shows negative direction (clockwise). 
The figure indicates that 
the azimuthal field reverses the direction with the radius.

Figure \ref{fig4} indicates that long wavelength magnetic loops 
are formed around disk-halo interface. 
Therefore, we discuss the formation mechanism of 
the magnetic loops. 
Horizontal magnetic fields embedded in gravitationally stratified 
atmosphere become unstable against long wavelength undular perturbations.
This undular mode of the magnetic buoyancy instability, 
called Parker instability,
creates buoyantly rising magnetic loops \citep{par1966}. 
Nonlinear growth of Parker instability in gravitationally stratified disks 
was studied by magneto-hydrodynamic simulations by \citet{mat1988}. 
They showed that magnetic loops continue to rise when $\beta < 5$.  
In weakly magnetized region where $\beta >5$, 
Parker instability only drives nonlinear oscillations \citep{mat1990}. 

In differentially rotating disks, MRI couples with Parker instability. 
The growth time scale of the non-axisymmetric MRI is $ t_{\rm MRI} \sim 1/(0.1\Omega) \sim 10 H/c_{\rm s}$ 
where $c_{\rm s}$ is the sound speed. On the other hand, the growth time scale of Parker instability 
is $t_{\rm PI} \sim 5H/v_{\rm A} \sim 5\sqrt{\beta \gamma/2} (H/c_{\rm s})$ where $v_{\rm A}$ and $\gamma$ are  
the Alfv\'en speed and the specific heat ratio, respectively. 
This time scale becomes comparable to the growth time of MRI 
when $\beta \sim 5$. 
Since magnetic turbulence produced by 
MRI limits the horizontal length of coherent magnetic fields, 
the growth of Parker instability can be suppressed in weakly magnetized 
region inside the disk.  
On the other hand, in  
regions where $\beta < 5$, 
nonlinear growth of Parker instability creates magnetic loops 
buoyantly rising to the disk corona \citep{mac2000, mac2009}.

Figure \ref{fig5}a shows an example of a magnetic loop at $t=1000t_0$. 
Magnetic loops are identified by the same algorithm as that reported in 
\citet{mac2009}. 
Figure \ref{fig5}b, \ref{fig5}c, and \ref{fig5}d show the distribution  
of the vertical velocity, plasma $\beta$, and density, respectively 
along the magnetic field line depicted in figure \ref{fig5}a. 
A black circle 
shows the starting point of the 
integration of a magnetic field line. 
The positive vertical velocity 
indicates that the magnetic loop is rising. 
The value of plasma $\beta$ 
decreases from the 
foot-point of the loop where $\beta \sim 100$ 
to the loop top where $\beta < 5$ (figure \ref{fig5}d). 
The density also decreases from the foot-points to the loop top.
These are typical structure of buoyantly rising magnetic loops 
formed by Parker instability.

From the numerical results presented so far, 
the initial magnetic field was found to be amplified and transported 
to the halo region after several rotation periods. 
In order to investigate the time evolution of the azimuthal magnetic field, 
we show a butterfly diagram of the azimuthal magnetic fields
at $\varpi/\varpi_0 = 2$ (figure \ref{fig6}a).  
The horizontal axis shows time $t/t_0$, and 
the vertical axis shows the height $z/\varpi_0$. 
Color denotes the direction of the azimuthal magnetic fields. 
White curve shows the iso-contour 
where $\beta = 5$ above the equatorial plane.
We should note that 
there is no magnetic flux 
at $\varpi/\varpi_0 = 2$ initially. 
As the time elapses, the accretion of the interstellar gas
driven by MRI transports magnetic fields from 
$\varpi/\varpi_0 = 10$ to $\varpi/\varpi_0= 2$.  
The amplified azimuthal magnetic fields change their directions quasi-periodically. 
Turbulent magnetic fields are dominant around the equatorial plane. 

When the magnetic tension becomes comparable to 
that of turbulent
motion in the disk surface
and coherent length of the 
magnetic field line increases, 
the magnetic flux buoyantly 
escapes from the disk due to Parker instability. 
We find that  the instability significantly grows,  
when the plasma $\beta$ decreases to around 5.  
The averaged escape velocity of the flux is about 25 km/s, 
approximately equal to the Alfv\'en velocity.

The periodic reversals of azimuthal magnetic fields can also be seen in 
correlation between the fields above and below the equatorial plane 
(figure \ref{fig6}(b)).
White color shows the region where $B_{\varphi}(z) \cdot B_{\varphi}(-z) > 0$,
and black shows the region where $B_{\varphi}(z) \cdot B_{\varphi}(-z) <0$. 
We can see fine structures of turbulence near the equatorial plane. 
In the halo region where $z/\varpi_0>1$, the color of 
outgoing flux changes quasi-periodically. 
It means that the disk magnetic topology changes between symmetric state and 
anti-symmetric state. 

Finally, we check the spatial distribution of the azimuthal magnetic fields.
The azimuthal magnetic fields in $\varpi- z$ plane averaged in the 
region $0 \leq \varphi \leq 2 \pi$ at $t/t_0 = 1000$is shown in figure \ref{fig7}a. 
We can see that a few $\mu {\rm G}$ fields are distributed as high as 
at $z/\varpi_0 = 5$.  
Figure \ref{fig7}b, and \ref{fig7}c show the distribution 
of the azimuthal magnetic fields in $\varpi-\varphi$ plane, 
where the azimuthal magnetic fields are averaged in the halo 
($1 <z/\varpi_0 < 1.5$) and in the disk ($0 <z/\varpi_0 < 0.5$), respectively. 
Since turbulent magnetic fields are dominant inside the disk,  
short magnetic filaments are formed (figure \ref{fig7}c ). 
Long and strong azimuthal magnetic sectors are formed in the halo region, 
because the plasma $\beta$ in the halo region becomes lower than 
that in the disk due to the density decrease (figure \ref{fig7}b). 
Therefore,   magnetic tension of buoyantly rising magnetic loop exceeds the ram pressure 
by turbulent motion around the disk surface and in the halo. 
These magnetic loops buoyantly escape from the disk by Parker instability 
and create the rising magnetic fluxes in the butterfly diagram shown in figure \ref{fig6}a.

\subsection{Dependence on the azimuthal resolution}

In order to study the dependence of numerical results on azimuthal resolution, 
we carried out simulations in which we used 64 (model ASYM64) and 
256 (model ASYM256) grid points in $0 \leq \varphi \leq 2 \pi$. 
Other parameters are the same as those in model ASYM. 

Figure \ref{fig8}a shows the time evolution of the magnetic energy 
averaged in the same region as figure \ref{fig1}c.  
Black, dashed and gray curves show model ASYM, ASYM256, and ASYM64, respectively. 
This panel indicates that the saturation level of the magnetic energy 
near the equatorial plane
slightly decreases as the resolution increases  
because shorter wavelength turbulence inside the disk 
can be resolved better in the model with higher azimuthal resolution, 
so that more magnetic energy dissipates in the nonlinear stage. 
The time evolution of the azimuthal magnetic fields averaged in the region 
where $5 < \varpi/\varpi_0 < 6$, $0 <z/\varpi_0 < 1$, and $0 \leq \varphi \leq 2\pi$
is shown in figure \ref{fig8}b. 
Colors are same as figure \ref{fig8}a. 
The amplitude of the azimuthal magnetic fields and the time scale of the field reversal 
are almost similar in model ASYM and ASYM256.   
In order to resolve the most unstable wavelength of MRI, 
we need more than 20 grid points per disk thickness for azimuthal direction
\citep{haw2011}.
Although ASYM does not satisfy this condition, 
the qualitative behavior of the results for ASYM is identical to that of ASYM256, 
which satisfies the condition (see figure \ref{fig8}b). 
For example, the ratio of the radial magnetic energy to the azimuthal magnetic energy  
($B_{\varpi}^2/B_{\varphi}^2$) is $0.12$ in model ASYM and $0.13$ in model ASYM256. 
Hence, we conclude that numerical results of model PSYM and ASYM in this paper 
obtained by simulations with 128 grid points in the azimuthal direction 
do not significantly differ from results with higher azimuthal resolution 
in the nonlinear stage.

\section{Summary and Discussion} 

We carried out 3D MHD simulations of the galactic gaseous disk. 
The numerical results indicate that magnetic fields are amplified by MRI. 
The amplified azimuthal magnetic fluxes 
buoyantly escape from the disk by Parker instability. 
Azimuthal component of the mean magnetic field reverses its  
direction quasi-periodically. 
The timescale of the reversal is about $300 t_0$ which corresponds to 
about 1.5 Gyr.

\begin{figure}
\plotone{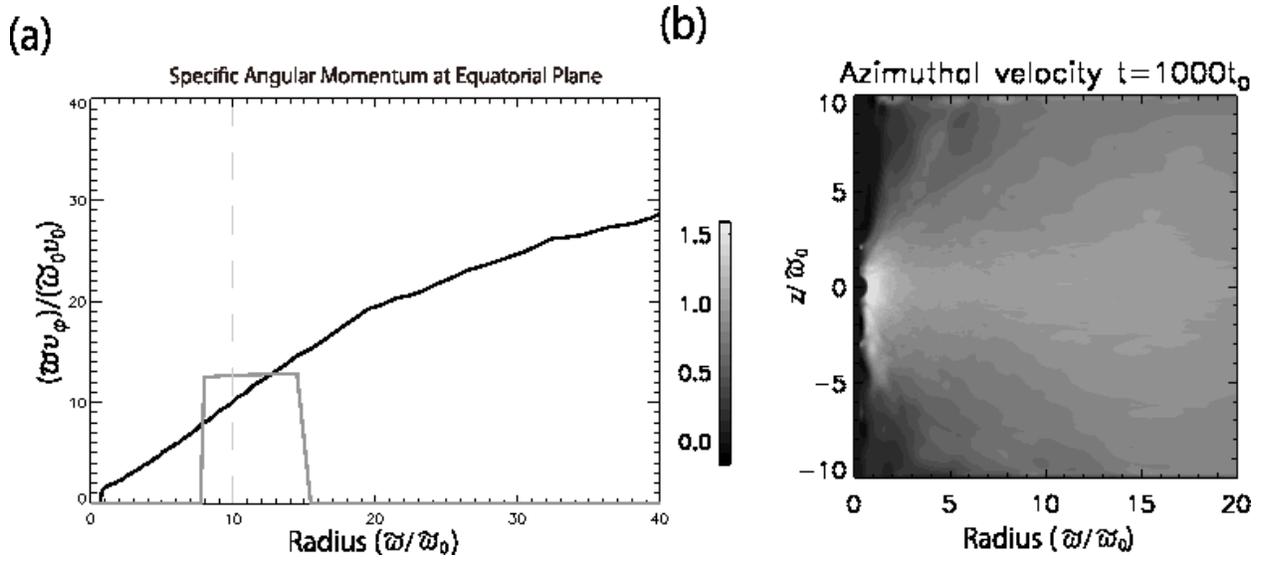}
\caption{ 
(a) Radial distribution of the specific angular momentum
($l= \varpi v_{\varphi}/(\varpi_0 v_0) $)
 at equatorial plane. 
Black curve shows  
the distribution at $t = 1000 t_0$, 
and gray curve shows the initial condition. 
Light gray dashed line shows the position of the initial density maximum. 
(b) Spatial distribution of the azimuthally averaged rotation speed at $t=1000 t_0$.
\label{fig3}}
\end{figure}

\begin{figure}
\plotone{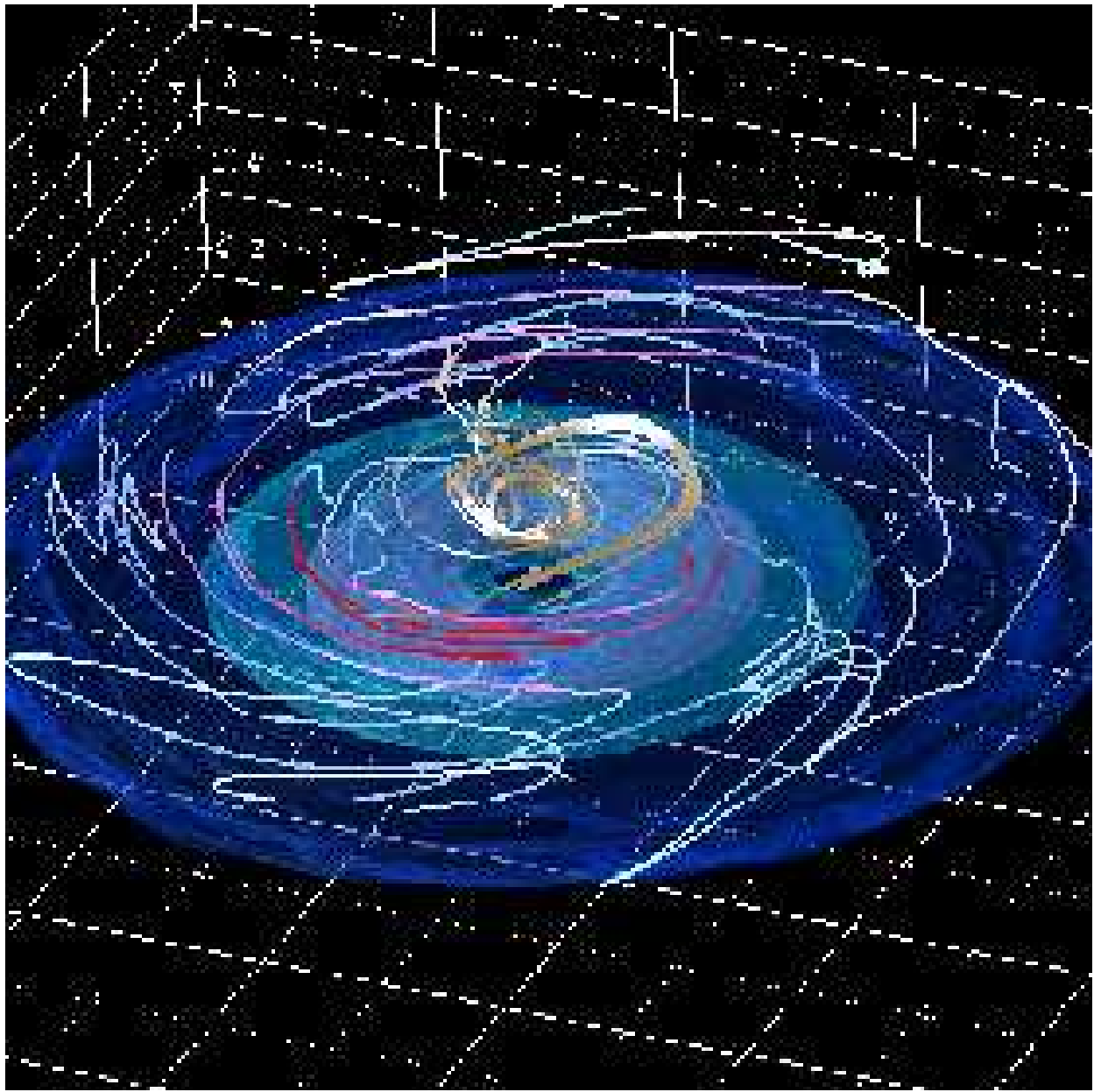}
\caption{Isosurface of density at $t=1000 t_0$.  
The region $|x|/\varpi_0, |y|/\varpi_0 < 18.5$, and $|z|/\varpi_0 < 7$ is plotted.  
Curves show the mean magnetic field lines. 
Color denotes direction of the azimuthal magnetic fields. 
Blue-white-red corresponds to negative-positive. 
\label{fig4}}
\end{figure}

\begin{figure}
\plotone{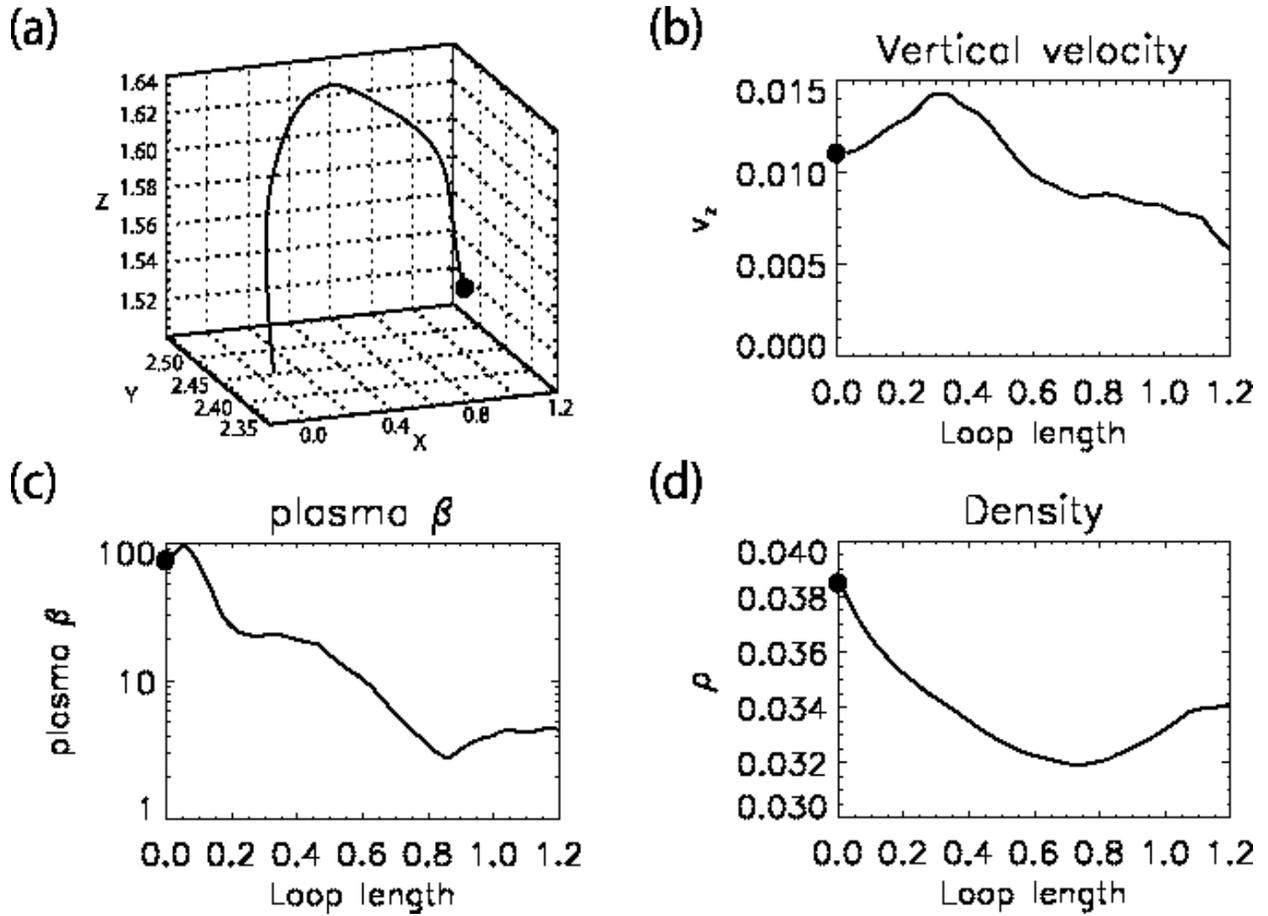}
\caption{(a) A typical magnetic loop at $t = 1000 t_0$. 
Black circle denotes the starting point of integration of the 
magnetic field line. 
The distribution of the physical quantities along the magnetic field line,  
(b) vertical velocity, (c) plasma $\beta$, and (d) density, respectively. 
\label{fig5}}
\end{figure}

The numerical results also show that 
galactic gaseous disk becomes turbulent 
and that averaged plasma $\beta$ 
stays at around $\beta \sim 10$. 
Plasma $\beta$ becomes lower near the disk surface due to the density stratification. 
Since rotational speed 
decreases from the disk to the halo
(see figure \ref{fig3}b), 
turbulent magnetic fields are stretched around the disk surface,   
forming ordered fields.
When the magnetic pressure of the ordered fields becomes 
as large as $\beta \sim 5$, 
magnetic flux escapes from the disk due to the buoyancy.

Figure \ref{fig9} schematically illustrates the mechanism 
of the galactic dynamo obtained from numerical simulations. 
Let us consider a differentially rotating disk 
threaded by weak seed fields,  
which rotates counter clockwise.  
As MRI grows in the gaseous disk, 
positive and negative fields whose strengths are comparable are amplified.   
When the averaged magnetic pressure becomes about 10\% of the gas pressure in the disk, 
 local magnetic pressure near the surface exceeds 20\% of the gas pressure and 
Parker instability creates magnetic loops  near the disk surface. 
Since the magnetic field strength parallel to the seed field grows 
faster than that of the opposite field,  
the counter clockwise fields satisfy 
the condition of Parker instability earlier 
and buoyantly escape from the disk.  
Subsequently, clockwise magnetic fields which remain inside the disk are amplified by MRI, 
and mean azimuthal magnetic fields inside the disk reverses their direction.  
The reversed fields become a seed field for the next cycle.  

Similar quasi-periodic reversal of azimuthal fields have been reported in 
local 3D MHD simulations of a differentially rotating disk
\citep{shi2010}.   
Although MRI can grow so long as  $\beta \geq 0.1$ \citep{joh2008}, 
the saturation level of plasma $\beta$ is about $\beta \sim 1- 10$ 
when the simulations include the effect of the vertical stratification. 
The growth of MRI saturates 
when the growth rate of the Parker instability is largest ($\beta \sim 1-10$).  
It indicates that the saturation levels of the magnetic energy inside the disk
is determined by the Parker instability around the surface layer. 

Our finding about the periodical reversal of the direction of the halo azimuthal 
field would impact on the observation of all sky RM map. 
There are a number of reports 
about all sky RM distributions \citep{tay2009, bra2010, han2007}. 
Study of all sky RM distribution is essential not only to understand the global 
structure of galactic magnetic fields 
but also to explore the intergalactic magnetic fields in the intergalactic medium 
(Akahori \& Ryu 2010, 2011, references therein). 
For instance, \citet{tay2009} suggested that the existence of the halo poloidal field 
 is consistent with a dipole field rather than a quadrupole field.

We calculated the distribution of RM obtained from numerical results (figure \ref{fig10}). 
The position of the observer is assumed to be at $r =8 {\rm kpc}$ at $t = 1000 t_0$.  
The direction of the rotational velocity and azimuthal magnetic fields 
is counter-clockwise in our numerical simulation. 
Since the angular velocity observed in Milky way is clockwise, 
we calculated the RM distribution  by reversing the $z$-direction 
(i.e. north is $-z$ direction).  
From the characteristics of MHD, if the initial magnetic field direction 
were reversed, we can obtain the same result except for the direction of magnetic fields. 
Thus we reversed the color of the RM from the original one.
The scale of RM is arbitrarily  in this paper. 
We found that magnetic loops produced by Parker instability 
formed multiple reversals of the the sign of RMs along the latitude. 
These reversals reveal that magnetic fields with opposite polarity 
emerge from the disk periodically. 
The snapshot shown in figure \ref{fig10} indicates that  
the distribution of RM is point symmetric with respect to the galactic center, 
and that this feature remains in timescale of $\sim 1.5$ Gyr (figure 2a). 

Braun et al. (2010) observed RM distributions in nearby galaxies. 
They also pointed out that the field topology in the upper halo of galaxies is 
a mixture of axi-symmetric spiral quadrupole field in thick disks  
and raidally directed dipole field in halos.
They suggested that the origin of the dipole components might be a bipolar outflow. 
Mixed structure was also suggested by recent studies of all-sky RM map 
(see, e.g., figure 8 of Pshirkov et al. 2011). 
The RM distribution in the halo region of our numerical results shows 
an anti-symmetric distribution corresponding to the dipole field.  
 
\begin{figure}
\plotone{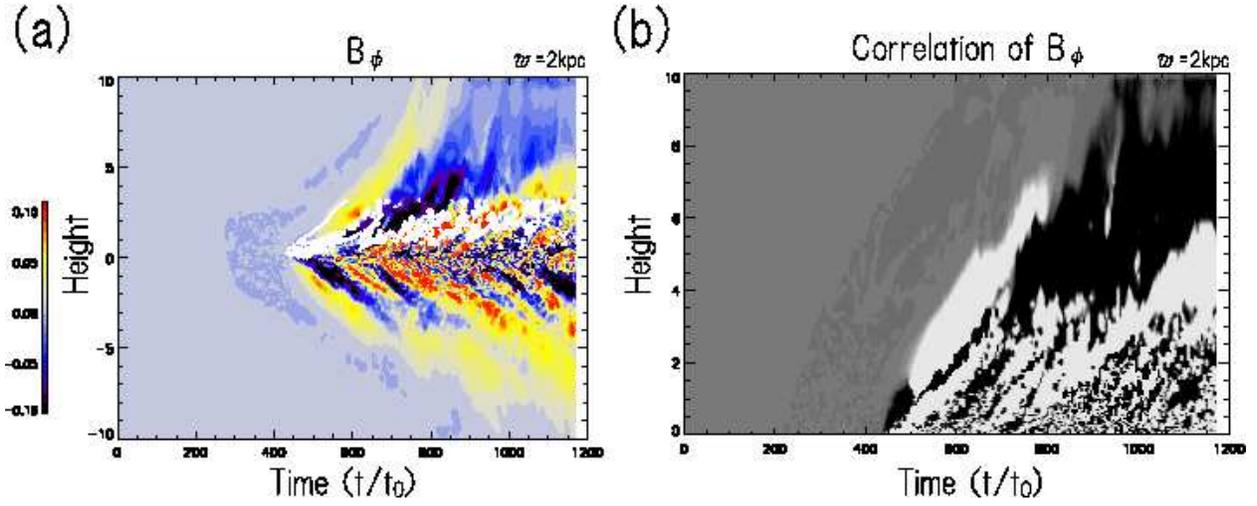}
\caption{(a) Time evolution of the azimuthal component of magnetic field 
averaged in the azimuthal direction. 
White curve shows the location where plasma $\beta = 5$ 
above the equatorial plane $(z > 0)$. 
(b) Correlation of the direction of $B_{\varphi}$ below the equatorial plane and 
above it. White shows 
the region of positive correlation where the direction of azimuthal magnetic fields 
above and below the disk is identical, 
and black is 
the region with negative correlation. 
\label{fig6}}
\end{figure}

\begin{figure}
\plotone{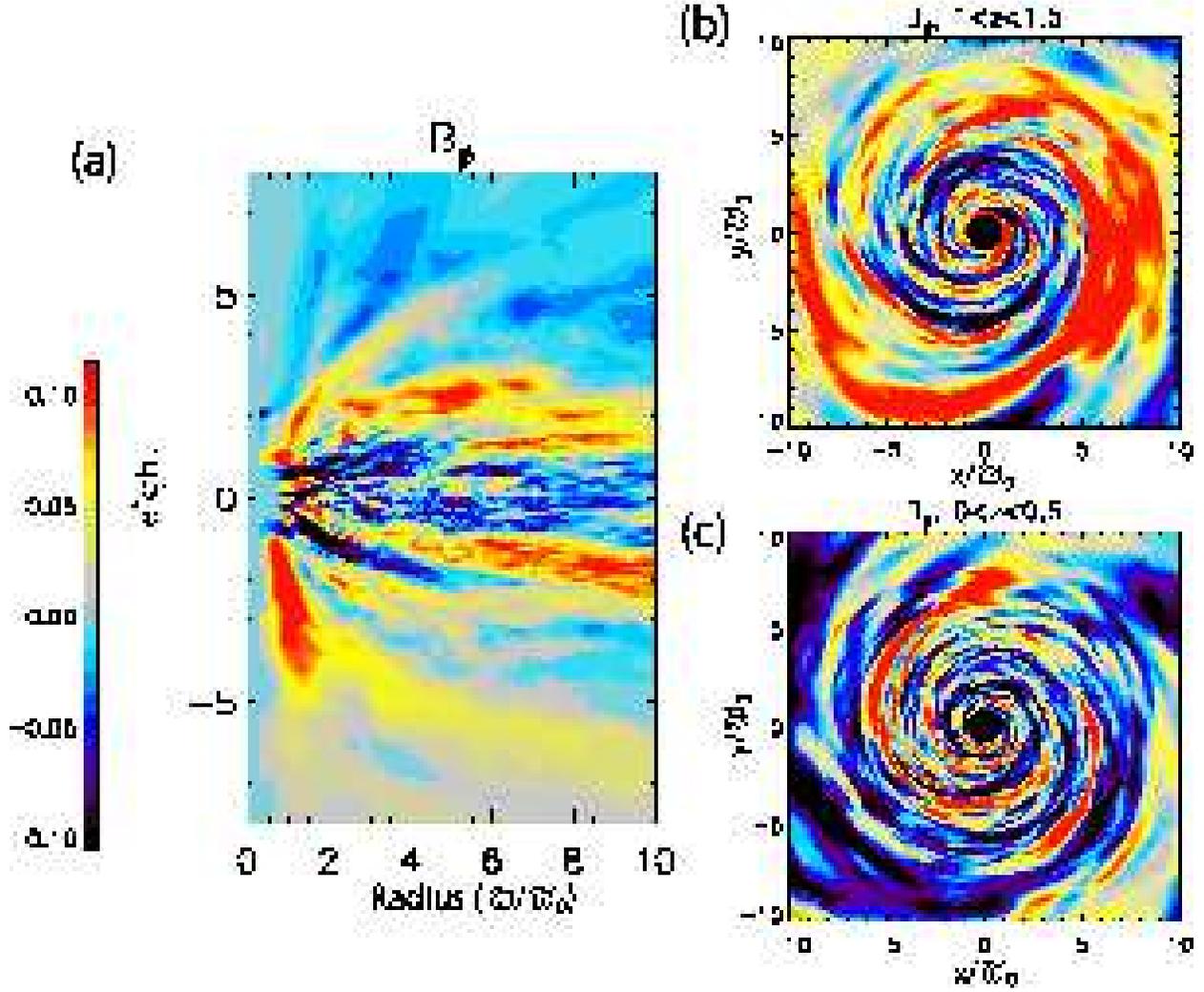}
\caption{(a)  $\varpi-z$
distribution of the azimuthal magnetic fields averaged in the 
azimuthal direction at $t/t_0 =1000$. (b) $\varpi-\varphi$ distribution of 
azimuthal magnetic fields
averaged in the vertical direction where $1 \leq z/\varpi_0 \leq 1.5$. 
(c) $\varpi-\varphi$ distribution of the azimuthal magnetic fields 
averaged in $0 \leq z/\varpi_0 \leq 0.5$. 
\label{fig7}}
\end{figure}

\begin{figure}
\plotone{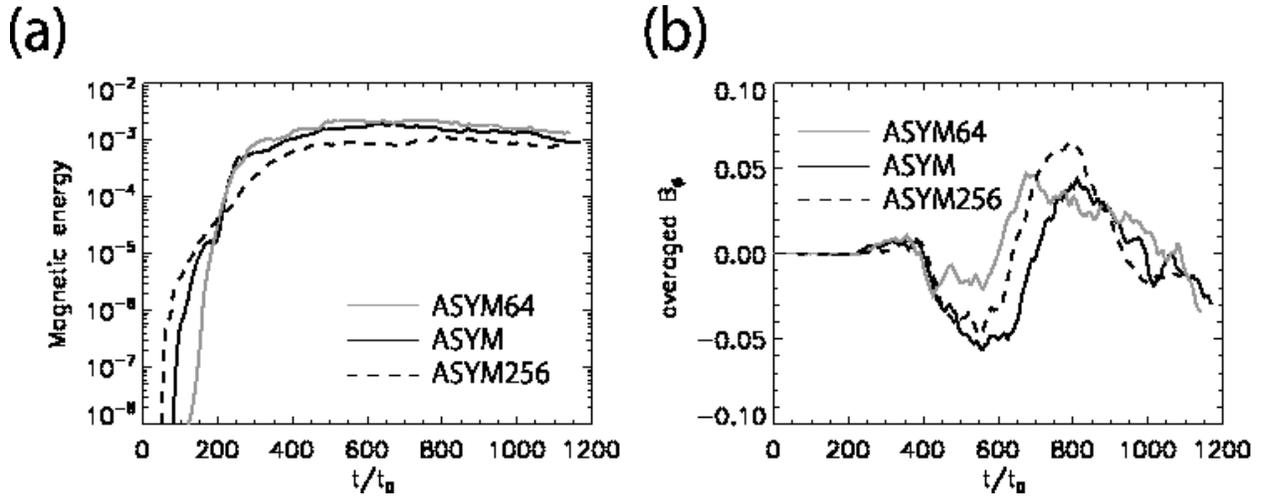}
\caption{(a) Time evolution of the magnetic energy 
averaged in the region where $ 2< \varpi/\varpi_0 <5$, 
$|z| /\varpi_0 < 1$, and $0 \leq \varphi \leq 2 \pi$ . 
(b) Time evolution of the azimuthal magnetic field averaged in the region where 
$5 <\varpi/\varpi_0 <6$, $0 <z/\varpi_0<1$, and $0 \leq \varphi \leq 2 \pi$. 
Black, gray, and dashed curves show the results for Model ASYM, 
ASYM64, and ASYM256, respectively.  
\label{fig8}}
\end{figure}

\begin{figure}
\plotone{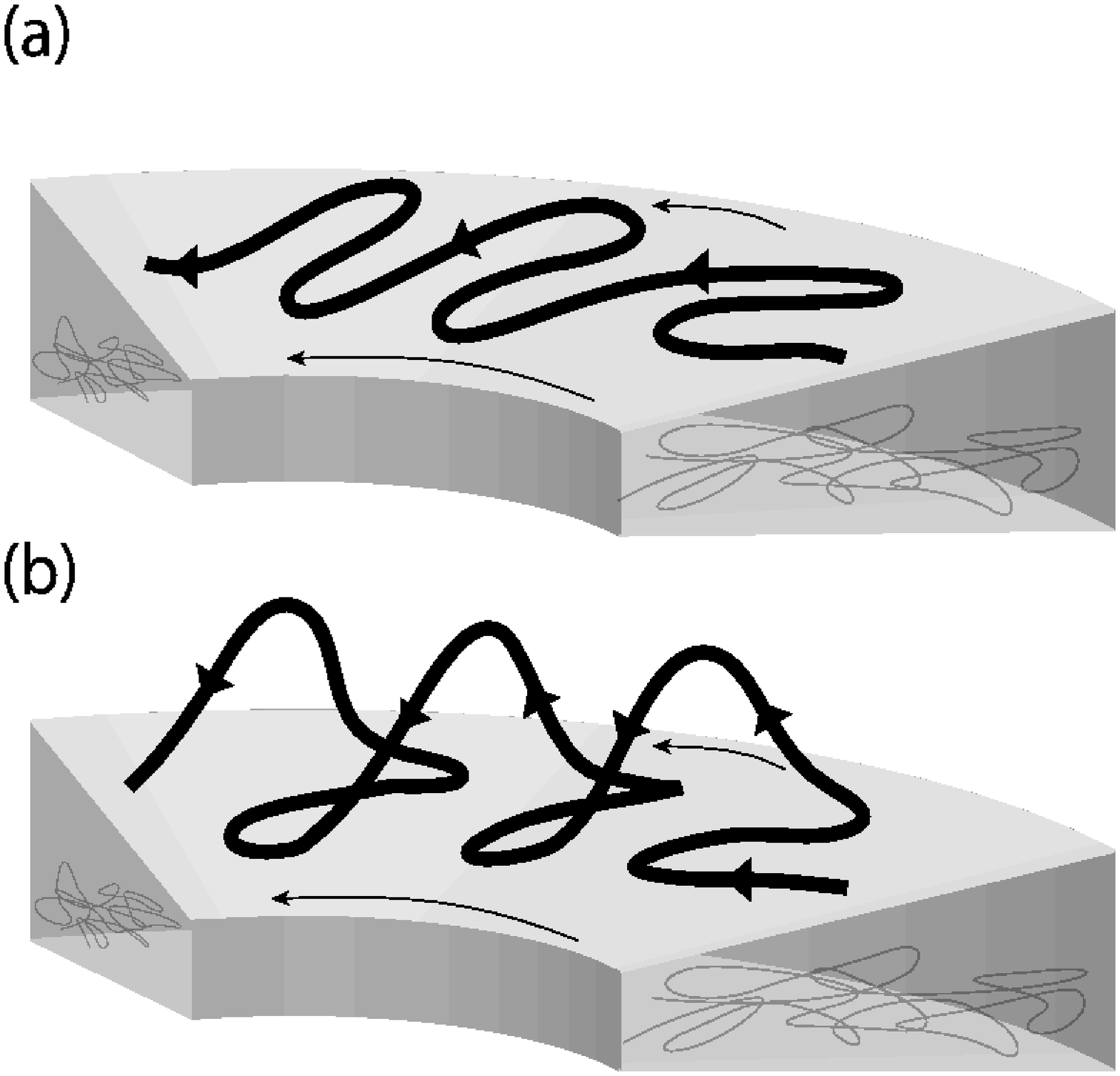}
\caption{Schematic drawing of the mechanism of MHD Dynamo.
\label{fig9}}
\end{figure}

\begin{figure}
\plotone{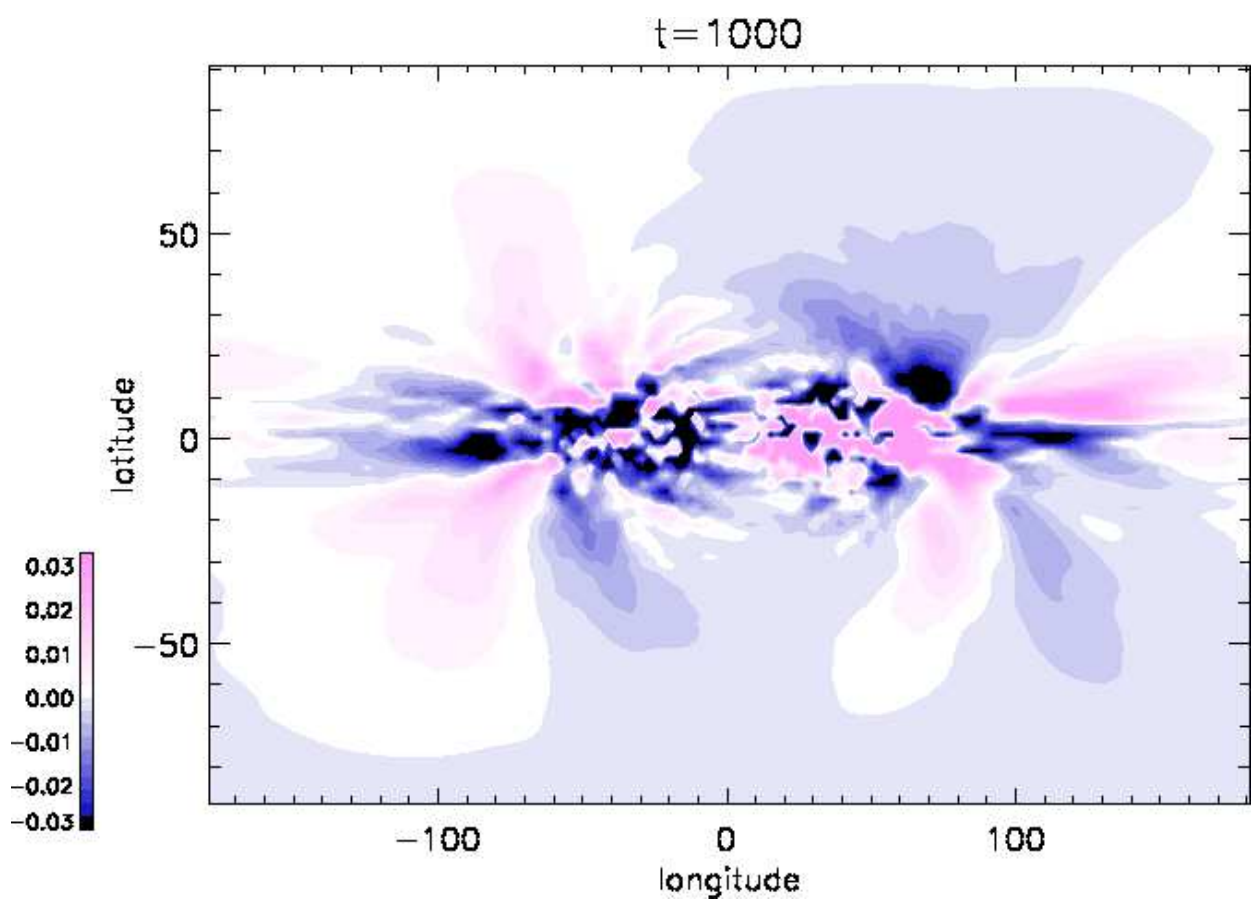}
\caption{Rotation measure distribution obtained from numerical results.  
\label{fig10}}
\end{figure}

Outflows are produced by emerging magnetic fluxs  driven by  
Parker instability. 
The time interval of the buoyant escape of the magnetic flux by Parker instability 
is about 10 rotational periods 
which corresponds to the growth timescale of MRI. 
Since buoyant escape of magnetic flux takes place about 10 cycles during the 
simulation for $4 {\rm Gyr}$
at $\varpi/\varpi_0 = 2$,   
the disk becomes turbulent enough, and 
the influence of the initial field directions become negligible. 
Even when we start simulations with magnetic fields symmetric to  
the equatorial plane,  
the direction of the halo fields in the upper and lower 
hemisphere becomes opposite at $\varpi/\varpi_0 = 2$ 
as shown in figure \ref{fig6}a. 
On the other hand, 
the effect of the initial field topology still remains in the outer disk region. 
Since RM is the sum of the line of sight value, 
the high latitude region 
(i.e., magnetic loops in nearby disks) still memorizes initial fields. 
Since the outflow velocity may be faster when we include the 
effect of the super nova explosions  \citep{wad2001}
and the cosmic ray pressure \citep{kuw2004, han2009}, 
outer disks may have enough time to create halo fields by the 
buoyant escape of turbulent disk fields, and  
eliminate the effect of the initial fields.

Our numerical calculation showed that 
initially non-rotating halo rotates after several rotation periods, 
because the angular momentum was supplied from the disk to halo 
by rising magnetic loops formed by the Parker instability.  
In fact, HI observation of late type spiral galaxy, NGC 6503 indicated that 
there were two components of HI disk, thin dense disk and thick low density disk, 
and that the  
thick low-density disk rotates slower than the thin disk \citep{gre2009}. 
The rotation speed of the ionized gas in the thick disk is slower than that of HI \citep{bot1989}.

In this paper, we assumed that  
the interstellar gas is one component gas with  
temperature $2 \times 10^5$ K in order to compare the result with 
that by \citet{nis2006}. 
However, 
in realistic galaxy, 
the interstellar gas has 
multi-temperature, multi-phase components. 
%
In future works, we would like to include the energy input by supernovae, 
multi-temperature structure of the interstellar gas, and cosmic-rays.

\acknowledgments
We are grateful to Drs. F. Cattaneo, R. Beck, K. Takahashi, and K. Ichiki 
for useful discussion. Numerical computations were carried out on SX-9 
at Center for Comupational Astrophysics, CfCA of NAOJ (P.I. MM). 
This work is financially supported in part by a Grant-in-Aid for 
Scientific Research (KAKENHI) from JSPS 
(P.I. MM:23740153, P.I. RM:23340040, 24111704, P.I. TK:23540274). 
This work is also supported by JSPS Core-to-Core program No. 22001 (RM).
T.A. acknowledges the supports of the Korea Research Council of Fundamental Science and Technology (KRCF) and the Japan Society for the Promotion of Science (JSPS).

\clearpage

\clearpage

\end{document}